\documentclass[aps,prb,twocolumn,groupedaddress,showpacs,showkeys,floatfix]{revtex4}
\usepackage{amsmath}
\usepackage{amsfonts}
\usepackage{amssymb}
\usepackage{graphicx}

\begin{document}


\title{Unusual conductance of polyyne-based molecular wires}

\author{\v Z.~Crljen}
\email[]{crljen@irb.hr}

\author{G.~Baranovi\' c}
\affiliation{R. Bo\v skovi\' c Institute, P.O. Box 180, 10002 Zagreb,
     Croatia}

\date{May 4, 2006}

\begin{abstract}
We report a full self-consistent {\it  ab initio} calculation of the 
current-voltage curve and the conductance of thiolate capped polyynes 
in contact with gold electrodes. We find the conductance of polyynes 
an order of magnitude larger compared with other conjugated oligomers.
The reason lies in the position of the Fermi level deep in the HOMO 
related resonance. With the conductance weakly dependent on the applied
bias and almost independent of the length of the molecular chain, 
polyynes appear as nearly perfect molecular wires.

\end{abstract}
\maketitle

The study of transport properties of single molecules have attracted a
significant attention because of their potential use in molecular 
electronic devices. One of the major classes of molecules considered in 
conductivity studies, primarily for their molecular 
wire behavior,\cite{Schumm,Davis,Magoga,ferrocene,Crljen} is conjugated oligomers. 
They have shown a number of useful nonlinear properties such as 
conductance switching and negative differential resistance.\cite{Chen,TW}
However, inspite of a number of interesting experiments\cite{ferrocene,exp} a molecule 
with good molecular wire properties has not yet been spotted.   

A useful molecular wire should provide a high and stable conductance over
a wide bias region and for various lengths of the molecules. 
A linear chain of carbon atoms with double bonds between neighboring 
atoms, usually referred to as cumulene, was proposed as an ideal molecular
wire\cite{Schumm} and the calculations of the conductance of 
cumulene connected to gold electrodes were reported.\cite{Lang,Brandbyge}
Lang and Avouris\cite{Lang} showed that  
the conductance of cumulene did not stay constant in the ballistic regime,
but rather oscillated between the constant
values characteristics of the odd and even number of atoms in the chain. 

In this paper we show an entirely different behavior of polyynes, another
form of the carbon atom chain.
Polyynes are simple and yet most intriguing of conjugated organic oligomers.
Only recently, have they been assembled up to decayne.\cite{Slepkov} 
Formed as a linear chain of carbon atom 
pairs $(CC)_{n}$, with alternating single and triple bonds, they are a unique,
truly one-dimensional, molecular system. Two $\pi$-electron systems 
of the sp-hybridized structure provide polyyne with approximately cylindrical 
electronic delocalization along the conjugated backbone. The electronic 
transport is therefore independent of the rotation around the single bond, which
is a limitation often present in other organic oligomers.\cite{TW} 

We have obtained the electronic structure and transport properties of a series
of polyynes up to octayne, connected to gold electrodes. 
The stability of polyyne with respect to single- and triple-bond
alternations was achieved by fixing the molecule at the ends with 
thiol bonds. In addition, the thiol capped polyynes make a strong chemisorption
bond onto the metallic electrodes. 
We found that they had more than an order of magnitude higher conductance when
compared with other conjugated oligomers. In contrast to 
the cumulenes they were not 
prone to oscillations in conductance with the length of the molecule.  
We also found that their conductance 
was very weakly dependent on the molecular length and almost constant in the 
wide range of  bias voltages.

A previous study, with a different theoretical approach, included polyynes up to 
triyne with Pd contacts of a different contact geometry which resulted in lower
conductance.\cite{Seminario}

In order to perform the first-principle quantum modeling of the electronic 
structure under nonequilibrium conditions and to calculate the current-voltage 
characteristics and the differential conductance of the system, a full 
self-consistent {\it ab initio} method, which includes portions of the electrodes, 
had to be used.\cite{semi} The calculations were carried out using a 
nonequilibrium Green functions technique based on density functional theory, 
as implemented in the TranSIESTA simulation package.\cite{TranS}  
The current through the contact was calculated using the 
Landauer-Buttiker formula,\cite{Datta}
$I(V_{b}) = G_{0}\int^{\mu_{R}}_{\mu_{L}}\, T(E,V_{b})\, dE$,
where $G_{0}=2e^{2}/h$ is the quantum unit of conductance and $T(E,V_{b})$ 
is the transmission probability for electrons incident at an energy E 
through the device under the potential bias $V_{b}$. 
The difference between the electrochemical 
potentials $\mu_{L/R}$, of the left and right electrodes,
respectively, is $\mu_{L}-\mu_{R}=eV_{b}$. The computational procedure used 
was described extensively elsewhere\cite{Crljen}. 

\begin{figure}[t]
\resizebox{0.85\columnwidth}{!}{
\includegraphics[clip=true]{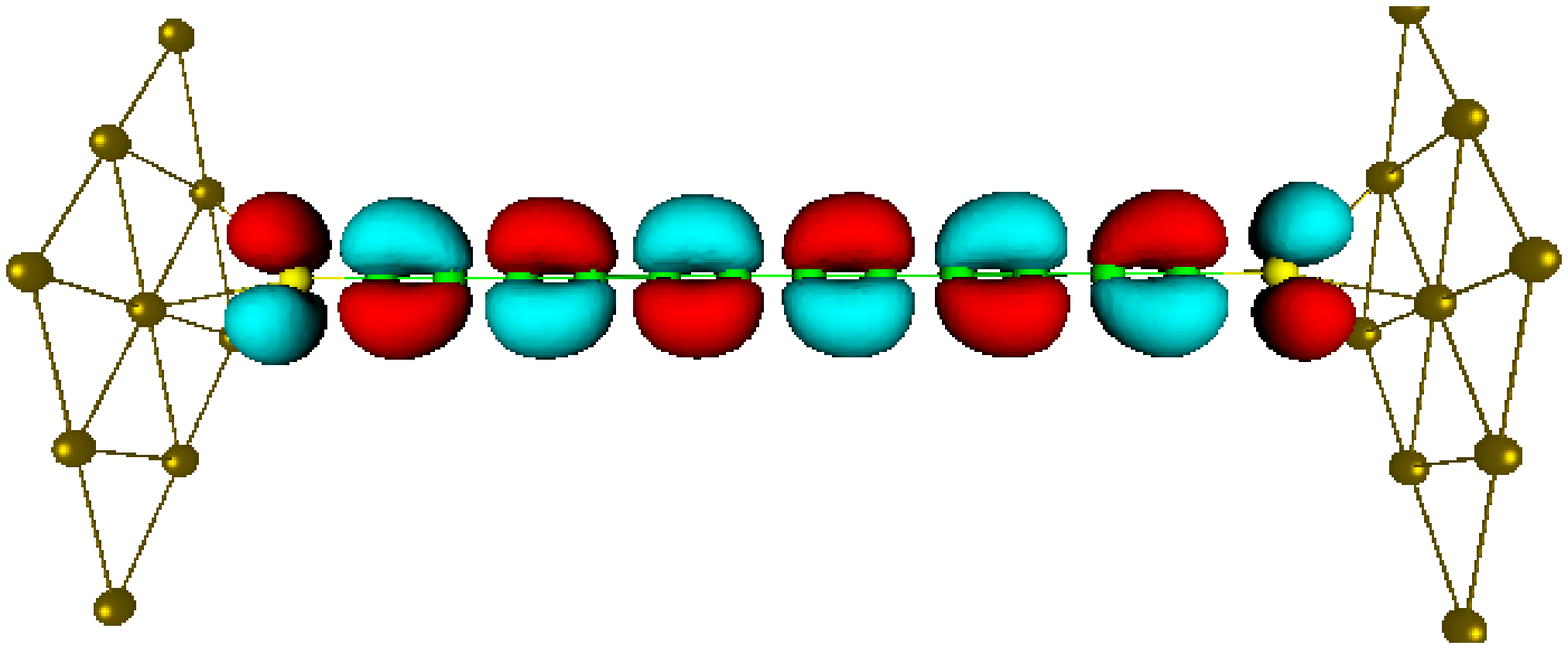}
}
\resizebox{0.85\columnwidth}{!}{
\includegraphics[clip=true]{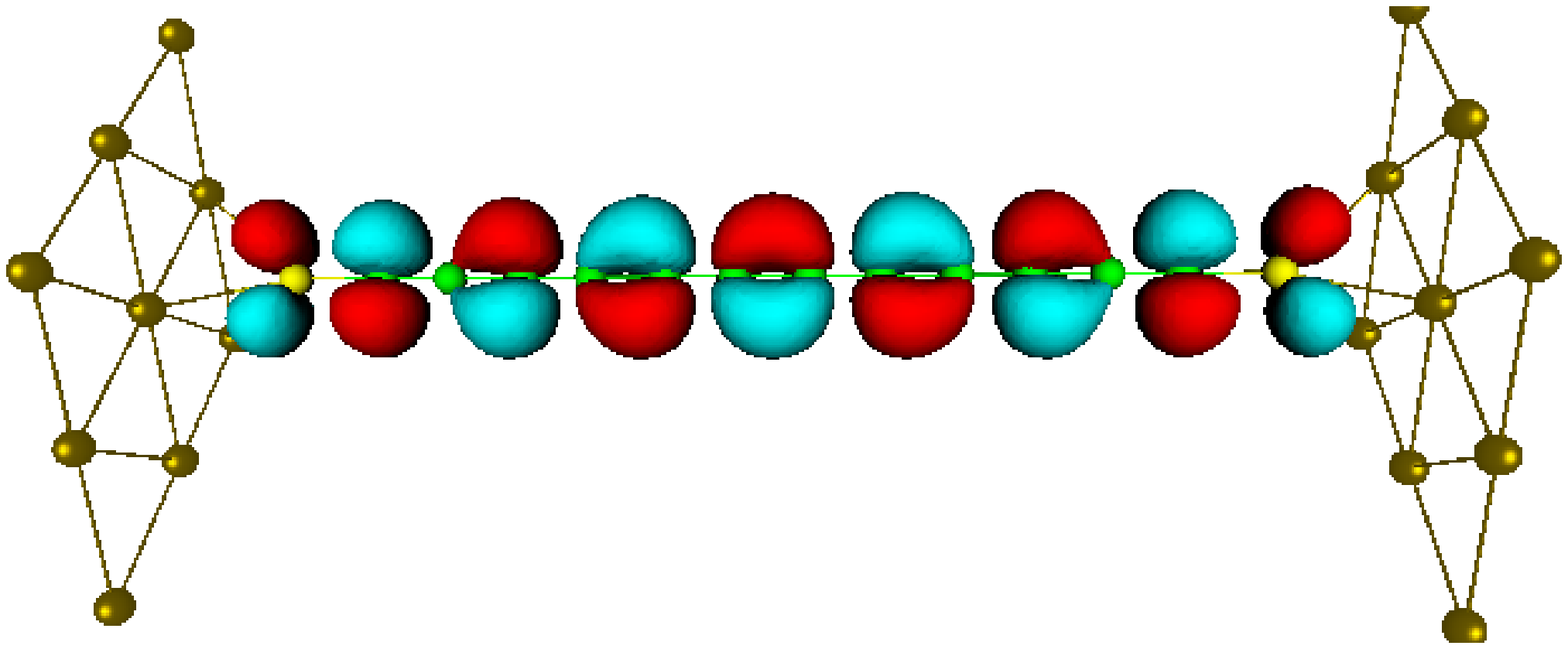}
}
\caption{\label{Au:hex}
Hexayne connected to two Au(111) surfaces via thiolate bonds, shown with 
the MPSH HOMO orbital (upper panel, notice the delocalized shape at
each side of single bond) and the MPSH LUMO orbital
(lower panel) at $V_{b}=0.6V$ bias voltage. 
}
\end{figure}

The molecular electronic system considered consisted of a monolayer of 
molecules coupled to two semi-infinite electrodes, as depicted 
in Fig.\ref{Au:hex} for hexayne. 
We optimized the geometry of free thiol capped polyyne in a separate DFT 
calculation\cite{optim}. The molecule was then positioned perpendicularly to
the z-direction in the hollow sites of both Au(111) surfaces of the electrodes
symmetrically at a favorable Au-S bonding distance\cite{cell}.
We let the molecular coordinate relax, keeping the gold atoms at their
bulk positions. A small change in the molecular geometry occurred with 
respect to the free molecule. We discuss this at the end of the paper.

\begin{figure}[t]
\resizebox{1.0\columnwidth}{!}{
\includegraphics[clip=true]{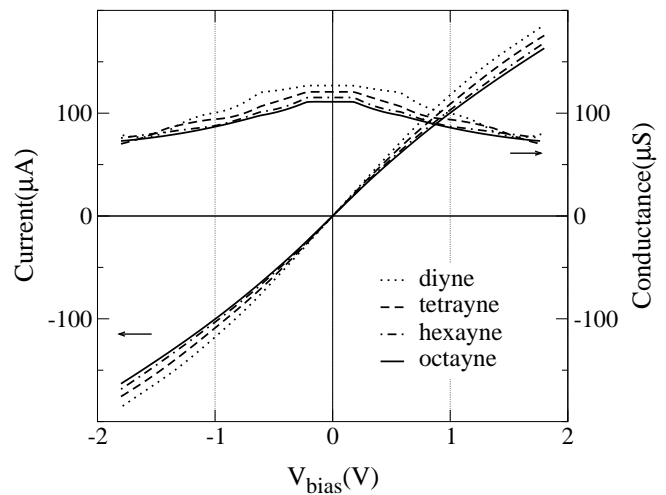}
}
\caption{\label{I-V_polyyne}
Current and differential conductance of polyyne systems as a function 
of bias voltage. A self-consistent calculation has been employed for each
bias voltage.  Notice the slow and smooth almost linear decrease of the conductance.
}
\end{figure}

The main characteristics of the obtained transmission spectra are the almost 
linear increase of the current with bias voltage and a high value of 
conductance over a wide bias region from -2V to 2V, as seen 
in Fig.\ref{I-V_polyyne}. 
The obtained spectra of all members\cite{others} of the polyyne series show 
a mutual pronounced similarity without oscillations with the length of 
the molecule.  
The conductance at zero bias voltage is
1.65 $G_{0}$ for diyne, 1.56 $G_{0}$ for tetrayne, 
1.49 $G_{0}$ for hexayne, and 1.44 $G_{0}$ for octayne. 
The overall high value of conductance is expected,
since the main transmission channels involve 
double degenerate molecular $\pi$ orbitals.
Unlike in the true ballistic transport in a quantum structure where a constant
conductance might be expected irrespective of the molecular 
length\cite{Landauer}, a slow decrease of conductance with the length of the
molecule is demonstrated. The decrease should be attributed to a weak
reduction of hybridization at the molecule-electrode contact with the number 
of atoms in the molecular chain\cite{width}. 

\begin{figure}[t]
\resizebox{0.95\columnwidth}{!}{
\includegraphics[clip=true]{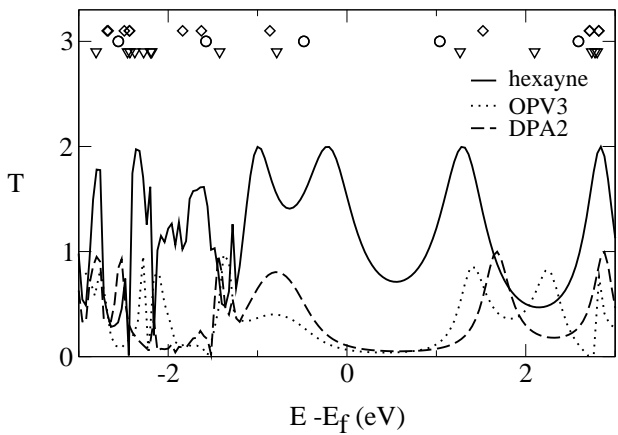}
}
\caption{\label{T_Vb0_all}
Transmission amplitude of hexayne compared with DPA2 and OPV3 at zero 
bias voltage. 
MPSH eigenvalues are marked with circles for hexayne, diamonds for DPA2, 
and triangles for OPV3.
}
\end{figure}

When compared with other conjugated oligomers, polyynes show much higher 
conductance. In Fig.\ref{T_Vb0_all} we compare the transmission amplitude
at zero bias of three different thiolate capped molecules: hexayne, 
diphenyl diacetylene (DPA2), and phenylene vinylene oligomer with three 
benzene rings (OPV3). They were all chemisorbed onto Au(111) electrodes 
in hollow positions at both ends in order to ensure the same bonding geometry
thus avoiding the possible bonding site effects on the considered transmission.
The resulting interelectrode separations for all three systems were rather 
similar, but the corresponding zero bias conductances differed by more than 
an order of magnitude, as shown in Table \ref{distance}. The large difference 
in conductance is a consequence of the entirely different electronic structure
and density of states at the Fermi level ($E_{f}$).   
The hybridization of the molecular level with the gold electrode states
broadens the level into the resonance. The width and the position of the 
resonance with respect to the Fermi level of the system depend on the 
internal structure of the molecule and its bonding to the electrode. 
In the hexayne case, the double degenerate molecular $\pi$ orbitals are 
involved. The resulting HOMO related resonance\cite{MPSH} is wide and shifted up 
in energy close to the Fermi level, as seen in Fig.\ref{T_Vb0_all}. In fact,  
the HOMO and LUMO related resonances strongly overlap giving a wide transmission
band with large density of states, which results in high transmission. 
The transmission of DPA2 and OPV3 systems is considerably smaller when 
compared with the hexayne, as shown in Fig.\ref{T_Vb0_all}. 
The reason for that is the lifted HOMO level degeneracy 
in DPA2 and OPV3 and the positioning of the HOMO level 
further below $E_{f}$, which results in the lower  density of 
states at the Fermi level.
\begin{table}[b]
\caption{\label{distance}
Interelectrode separation $d_{elec}$ and zero-bias conductance G($\mu$,$0$) 
of polyyne systems compared with molecules of similar length.
$d_{elec}$ is the distance between the surface gold planes of the 
electrodes. 
}
\begin{ruledtabular}
\begin{tabular}{cccc}
  &   OPV3 & DPA2 & hexayne  \\ \hline
      \cline{2-4}
$d_{elec} (nm)$ & $2.3$ & $1.95$ & $2.127$\\
G($\mu$,$0$) $(\mu S)$ &  2.1  & 6.12  & 111.1
\end{tabular}
\end{ruledtabular}
\end{table}

\begin{figure}[t]
\resizebox{0.95\columnwidth}{!}{
\includegraphics[clip=true]{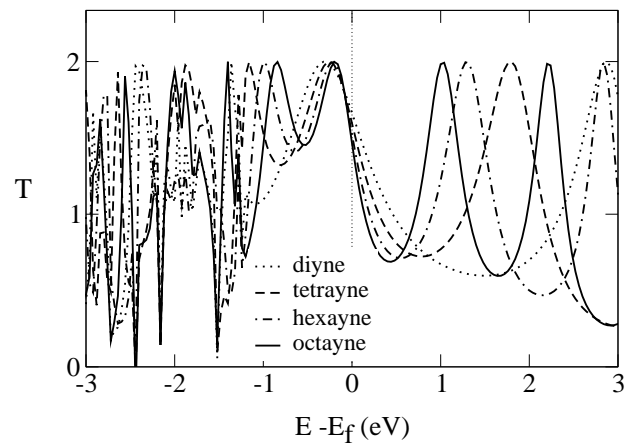}
}
\caption{\label{T_Vb0_polyyne}
Transmission amplitude of polyynes at zero 
bias voltage. Note the rather stable position of HOMO related peaks for all
polyynes.
}
\end{figure}

The high density of states at the Fermi level is the property of other 
members of the polyyne family, as well. 
With the increased length of the polyyne chain, levels are becoming more
densely distributed and of shorter width. The remarkable property is,
however, that the position of the HOMO related 
resonance moves slightly closer to $E_{f}$, thus compensating for the 
decrease of the density of states at $E_{f}$ due to the level 
sharpness, as seen in Fig.\ref{T_Vb0_polyyne}. That results in a very weak 
dependence of the zero bias conductance on the molecular length.

\begin{figure}[b]
\resizebox{0.95\columnwidth}{!}{
\includegraphics[clip=true]{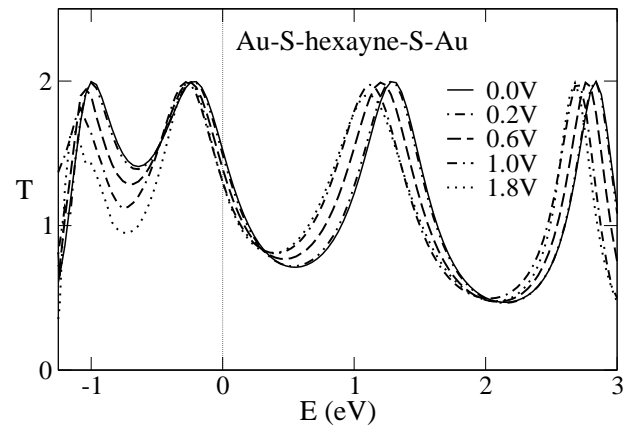}
}
\caption{\label{hexayne-bias}
Transmission amplitude of dithiolhexayne connected to gold electrodes as 
a function of bias voltage. Energies measured from the average 
electrochemical potential of the electrodes.
}
\end{figure}

The conductance of polyynes decreases smoothly with the increase of the 
bias voltage, as shown in Fig.\ref{I-V_polyyne}. In order to elucidate this behavior, we show the transmission amplitudes of hexayne 
for a set of bias values in Fig.\ref{hexayne-bias}.
A remarkable similarity, in both shape and value, of transmission followed 
by a slight shift of the position of LUMO and higher resonances 
with the bias is seen. The separation between HOMO and LUMO levels decreases
only by 0.11 eV at the bias of 1.8 V
with respect to the zero bias value. 
The consequence of the similarity of transmission spectra is the relatively 
weak dependence of the conductance of hexayne on the bias, 
as seen in Fig.\ref{I-V_polyyne}.
 
The other polyynes showed the dependence of transmission amplitudes when 
subjected to different bias voltages similar to those for hexayne.
With the change of the bias from 0 V to 2.0 V, the HOMO-LUMO
energy separation for all 
polyynes that we consider decreases by no more than 0.15 eV.
Such a small change in the molecular level positions relative the average
electrochemical potential of the electrodes ensures a slow decrease of
the conductance over the
entire bias region, as clearly seen in Fig.\ref{I-V_polyyne}.
 
What makes polyynes so much different from other conjugated polymeres is the
high density of states at the Fermi level. Even in a free molecule of 
polyyne, where the alternate single and triple bonds open up a HOMO-LUMO gap,
the strong polarizability and hyperpolarizability\cite{Slepkov, hyperpol} evolve,
owing to the large number of electrons. 
When thiol capped polyynes are chemisorbed onto gold electrodes, the strong 
hybridization of molecular states with the metallic 
electrode states results in a new electronic structure of the combined system.
The change of the level positions is so strong that the Fermi level 
of the system enters deep into the HOMO related resonance resulting 
in the pronounced metallic character of the system.

As already said, we used LDA for the exchange-correlation functional in our DFT 
calculations. There has been a discussion in the literature on the validity 
of DFT with the local density (LDA) and generalized gradient (GGA) 
approximations for $\pi$-conjugated systems\cite{locality}. 
Here, however, the $\pi$-conjugated system is connected to the electrodes 
and shows more metalliclike character. LDA- and GGA-based 
DFT calculations are therefore expected 
to give a respectable accuracy usually obtained in DFT calculations. 
The crucial parameters determining the HOMO-LUMO gap are the 
carbon-carbon bond lengths and their alternations.
We performed the calculations for isolated molecules of thiolate capped 
polyynes using the Gaussian03 program\cite{pot} with several XC functionals
and found that the GGA results did not substantially improve the values 
of the gap, nor the single and triple bond
lengths. Indeed, the difference in the HOMO-LUMO gaps was less 
than 0.037 eV between the PW91 (GGA) and  SVWN (LDA) based calculations, 
as seen in Table \ref{gap}. When extrapolated to infinite chain length the gap
stayed open, in agreement with the calculations with more exact exchange 
potentials of Weimer et al.\cite{Weimer}
For the molecule bonded to the electrodes, the  
HOMO-LUMO MPSH gap (first column of Table \ref{gap}) was reduced 
with respect to the isolated molecule as expected.
The reduction was slightly more pronounced for shorter molecules, equal to 
0.031 eV for diyne,
and decreased to 0.025 eV for octayne. Evidently, the HOMO-LUMO MPSH gap 
closely followed the gap of the isolated molecule.  
The possible underestimation of the gap in the LDA-based 
calculations compared with the nonlocal XC potential calculations would have no
major effect. The metalliclike character of the conductance appears to be the 
consequence of the hybridization of the $\pi$-electron systems of polyynes with the 
gold electrode states, irrespective of the details of the XC potential.
\begin{table}[t]
\caption{\label{gap}
MPSH HOMO-LUMO energy gap $\Delta E$, in $eV$, at zero bias of thiolate 
capped polyynes with Au electrodes compared with the HOMO-LUMO gaps of bare 
thiol capped polyynes (DFT calculations with SVWN, 
PW91, and PBE XC functionals).
}
\begin{ruledtabular}
\begin{tabular}{ccccccc}
 &   \multicolumn{3}{c} { $AuS(CC)_{n}SAu $}  & \multicolumn{3}{c} 
 {$HS(CC)_{n}SH$ }  \\  \cline{2-4} \cline{5-7}
n    &   $\Delta E$  & $E_{HOMO}$ & $E_{LUMO}$  & $\Delta E_{SVWN}$ & $\Delta E_{PW91}$  & $\Delta E_{PBE}$  \\  \hline  \cline{2-7}
$2$ &  $3.254$ & $-0.897$ & $2.357$ & $3.529$  & $3.559$  & $3.549$ \\
$4$ &  $2.063$ & $-0.643$ & $1.420$ & $2.333$  & $2.370$  & $2.364$ \\
$6$ &  $1.520$ & $-0.482$ & $1.038$ & $1.762$  & $1.797$  & $1.788$ \\
$8$ &  $1.211$ & $-0.385$ & $0.825$ & $1.428$  & $1.463$  & $1.458$ \\
\end{tabular}
\end{ruledtabular}
\end{table}

The hybridization with the gold states did not affect the bond alternation
in the electrode-connected hexayne, as seen in Table \ref{bond}. 
The slight reduction of single bonds and the expansion of triple bonds 
of the connected hexayne are noticed.
The change of the bond length is the interplay of the hybridization of the 
molecular states with the gold electrode states and the relaxation of 
position of molecular atoms. The amount of change is smaller when the
molecule is longer owing to the redistribution of 
the relaxation over the entire molecule.
\begin{table}[h]
\caption{\label{bond}
Average single and triple related bond distances, measured in $\AA$, of thiolate 
capped hexayne in contact with Au electrodes
compared  with the bare thiol capped hexayne bond distances.
}
\begin{ruledtabular}
\begin{tabular}{ccccc}
 &   $AuS(CC)_{6}SAu$  & \multicolumn{3}{c} { $HS(CC)_{6}SH$ }  \\  \cline{3-5}
 &               & $SVWN$ & $PW91$  & $PBE$  \\  \hline  \cline{2-5}
$d_{single}$ (\AA) &  $1.316$ & $1.321$  & $1.330$  & $1.332$ \\
$d_{triple}$ (\AA) &  $1.286$ & $1.249$  & $1.252$  & $1.253$ \\
\end{tabular}
\end{ruledtabular}
\end{table}

In this paper, we have shown that polyynes are molecules with high 
conductance over the large bias region 
almost irrespective of the length of the molecule considered. Although polyynes 
form $\pi$-conjugated systems, they show an almost metalliclike character
of the current transport, when connected to the electrodes.
This makes them a possible candidate 
for a good molecular wire to be considered in molecular nanoelectronics.

This work was supported in part by the Ministry of Science and Technology
of the Republic of Croatia.

\begin{thebibliography}{00}
\bibitem{optim}
The geometry of the molecules in the junction is assumed to be the same as the
isolated molecules, which are obtained at the level of SVWN/6-31+G*;
see Ref.~\onlinecite{pot}.
\bibitem{cell}
The electrode is modeled as a 3x3 unit cell, and the S atoms were placed at 1.87 $\AA$
from the fcc position of Au(111). The cell was big enough to avoid the cross linking of 
molecules.
\bibitem{others}
The results for triyne, pentayne, and heptayne (not shown here) fit the curves 
for even members of the series.
\bibitem{width}
The effect of the enhanced DOS at $E_{f}$ due to the small interelectrode
 separation for shorter molecules should also be present here;
 see Ref.~\onlinecite{Crljen}.
\bibitem{locality} 
The results of such schemes for polarizability and hyperpolarizability of 
isolated molecules were largely overestimated when compared with methods with 
more accurate nonlocal XC potentials. See Ref.~\onlinecite{hyperpol}.
\end{thebibliography}

\end{document}